\documentclass[aps,twocolumn,prd,superscriptaddress,floatfix,nofootinbib]{revtex4-1}
\usepackage{graphicx,natbib}
\usepackage{tabularx,ragged2e,booktabs,caption,amsmath}

\begin{document}

\renewcommand{\theequation}{\thesection.\arabic{equation}}
\newcommand{\re}{\mathop{\mathrm{Re}}}
\newcommand{\be}{\begin{equation}}
\newcommand{\ee}{\end{equation}}
\newcommand{\bea}{\begin{eqnarray}}
\newcommand{\eea}{\end{eqnarray}}
\title{Variations of the fine-structure constant $\alpha$ in exotic singularity models}

\author{Mariusz P. D\c{a}browski}
\email[]{mpdabfz@wmf.univ.szczecin.pl}
\affiliation{Institute of Physics, University of Szczecin, Wielkopolska 15, 70-451 Szczecin, Poland}
\affiliation{Copernicus Center for Interdisciplinary Studies, S{\l }awkowska 17, 31-016 Krak\'ow, Poland}
\author{Tomasz Denkiewicz}
\email[]{atomekd@wmf.univ.szczecin.pl}
\affiliation{Institute of Physics, University of Szczecin, Wielkopolska 15, 70-451 Szczecin, Poland}
\affiliation{Copernicus Center for Interdisciplinary Studies, S{\l }awkowska 17, 31-016 Krak\'ow, Poland}
\author{C. J. A. P. Martins}
\email[]{Carlos.Martins@astro.up.pt}
\affiliation{Centro de Astrofisica da Universidade do Porto, Rua das Estrelas, 4150-762 Porto, Portugal}
\author{P. E. Vielzeuf}
\email[]{Pauline.Vielzeuf@astro.up.pt}
\affiliation{Centro de Astrofisica da Universidade do Porto, Rua das Estrelas, 4150-762 Porto, Portugal}
\affiliation{Faculdade de Ci\^encias, Universidade do Porto, Rua do Campo Alegre 687, 4169-007 Porto, Portugal}
\begin{abstract}
Various classes of exotic singularity models have been studied as possible mimic models for the observed recent acceleration of the universe. Here we further study one of these classes and, under the assumption that they are phenomenological toy models for the behavior of an underlying scalar field which also couples to the electromagnetic sector of the theory, obtain the corresponding behavior of the fine-structure constant $\alpha$ for particular choices of model parameters that have been previously shown to be in reasonable agreement with cosmological observations. We then compare this predicted behavior with available measurements of $\alpha$, thus constraining this putative coupling to electromagnetism. We find that values of the coupling which would provide a good fit to spectroscopic measurements of $\alpha$ are in more than three-sigma tension with local atomic clock bounds. Future measurements by ESPRESSO and ELT-HIRES will provide a definitive test of these models.
\end{abstract}


\maketitle

\section{Introduction}

The discovery of cosmic acceleration from supernova observations \cite{SN1,SN2}, unveiled the presence of an unknown source of energy which can be modeled in the easiest approach by a cosmological constant $\Lambda$, resulting in the standard $\Lambda$CDM model. Despite the fact that a range of observational tests appears to be in good agreement with this model, the physical interpretation of $\Lambda$ remains ambiguous. Thus a range of alternative scenarios gradually emerged, the most natural of which ascribes dark energy to the presence of a dynamical scalar field. These alternatives have to be tested by the local and global cosmological observations.

One specific class of models aiming to mimic the observed dark energy behavior are the so-called exotic singularity models \cite{nojiri,AIP2010}. In fact, the emergence of exotic singularities is related to some physical fields which phenomenologically are mimicked in the form of a specific parametrization of the evolution of the scale factor. In other words, exotic singularity models may be seen as a toy-model parametrization of the evolution of a physical degree of freedom, such as a dynamical scalar field and its coupling to gravity and other fields.

The issue of exotic singularities in cosmology was investigated more intensively soon after the discovery of cosmic acceleration and the first example of such a singularity was a big-rip due to the non-canonical scalar field known as phantom \cite{phantom}. Then, other options such as a sudden future singularity (SFS) \cite{Varun,barrow04}, finite scale factor singularity (FSFS) \cite{nojiri,matzner}, a big separation \cite{nojiri}, and w-singularity \cite{wsing} and many others have been proposed (for a recent review see \cite{sesto2013}). These singularities are weak in the sense that both particles and extended objects can pass through them \cite{LFJ,adam}. It also emerged that models which contain these singularities can, with suitable parameter choices, fit current observations \cite{DHD,GHDD,DDGH,laszlo,Denkiewicz1,Balcerzak,f(rho)}.

Whenever dynamical scalar fields are present, one naturally expects them to couple to the rest of the model, unless a yet-unknown symmetry suppresses these couplings. In particular, a coupling to the electromagnetic sector will lead to spacetime variations of the fine-structure constant---see \cite{uzanLR} for a recent review. In fact there is some recent evidence for such a variation \cite{webb}, which a dedicated VLT/UVES Large Program is aiming to test \cite{LP1}. In any case, these spectroscopic measurements can be used as additional tests of the underlying theories, in particular if one makes the 'minimal' assumption that the same dynamical degree of freedom is responsible for the dark energy and the $\alpha$ variations \cite{nunes0,nunes}. This is the approach we will take here, though note that alternatives also exist, as discussed in \cite{pauline1,calabrese2}.

Thus if one envisages exotic singularity models as toy model parametrizations for an underlying dynamical scalar field, one may ask what variations of $\alpha$ will ensue. As shown in \cite{calabrese1,pauline1}, with the above minimal assumption this question can be answered without explicit knowledge of the field dynamics---the evolution of the dark energy equation of state and density are sufficient. (Additionally, there will be a parameter describing the strength of the coupling to electromagnetism, that is the evolution of the gauge kinetic function.) Thus we will consider some representative exotic singularity models which were shown (in our recent  \cite{lastpaper} and references therein) to be in reasonable agreement with current background cosmological data, and study the behavior of $\alpha$ therein, under the assumptions stated above.

The paper is organized as follows. In section \ref{sec:singu} we present a brief review of the exotic singularity models useful for our further study. In section \ref{sec:varalpha} we discuss the physics behind the variation of the fundamental constants and our specific assumptions regarding this class of models. The result of applying these to our study-case exotic singularity models will be exposed in section \ref{sec:results}. Our conclusions are given in section \ref{sec:conclusions}.

\section{Exotic singularity model phenomenology}\label{sec:singu}

In this section we will briefly review the phenomenology of some previously studied exotic singularity models that are in reasonable agreement with cosmological observations. While several classes of such singularities can be studied, we will be focusing here on SFS models. We will also briefly contrast these with a related alternative (FSFS models) which turn out not to provide observationally viable $\alpha$ models.

In these models one assumes the standard Einstein-Friedmann standard field equations for the energy density and pressure:
\begin{eqnarray}
\label{rho}
\rho(t)=\frac{3}{8\pi G}\left(\frac{\dot{a}^2}{a^2}+\frac{kc^2}{a^2}\right)\\
\label{p}
p(t)=-\frac{c^2}{8\pi G}\left(2\frac{\ddot{a}}{a}+\frac{\dot{a}^2}{a^2}+\frac{kc^2}{a^2}\right)
\end{eqnarray}
appended by the continuity equation:
\begin{equation}
\dot\rho(t)=-3\frac{\dot{a}}{a} \left[\rho(t)+\frac{p(t)}{c^2}\right]~~,
\end{equation}
where $a \equiv a(t)$ is the scale factor, the dot means the derivative with respect to physical time $t$, $G$ is the gravitational constant, $c$ is the speed of light, and the curvature index $k=0, \pm 1$. For further analysis we will set $k=0$, in agreement with observational results. The main assumption of these models resides in the scale factor which is parametrized differently than for the standard model and can be expressed as a function of the four parameters: $\delta$, $m$, $n$, $t_s$, namely
\begin{equation}
\label{a(t)}
a(t)=a_s \left[ \delta+(1-\delta)\left(\frac{t}{t_s}\right)^m-\delta\left(1-\frac{t}{t_s}\right)^n \right]~~.
\end{equation}
The parameter $m$ characterizes the evolution of the universe near the initial big-bang singularity at $t=0$, the parameter $\delta$ gives the standard Friedmann limit $\delta \to 0$, the parameter $n$ characterizes an exotic singularity (an SFS singularity appears for $1<n<2$ and an FSFS singularity appears for $0<n<1$), the parameter $t_s$ tells us the moment of an exotic singularity to appear during the evolution, and $a_s \equiv a(t_s)$.  The ansatz (\ref{a(t)}) is fully equivalent to the one applied in Ref. \cite{barrow04} but differs from that one proposed in Ref. \cite{JCAP13} which uses an exponential function of time. From the relation (\ref{a(t)}) one defines the redshift of an object being at radial distance $r_1$ at the moment $t_1$ with respect to an observer receiving the signal at $t_0$:
\be
\label{redshift}
1+z=\frac{a(t_0)}{a(t_1)} = \frac{\delta +
\left(1 - \delta \right) \left(\frac{t_0}{t_s}\right)^m - \delta \left( 1 - \frac{t_0}{t_s} \right)^n}
{\delta + \left(1 - \delta \right) \left(\frac{t_1}{t_s}\right)^m - \delta \left( 1 - \frac{t_1}{t_s}
\right)^n}~,
\ee
as well as the Hubble function
\begin{equation}
\label{eq:Hsing}
H(t(z))= \frac{1}{t_s}\frac{m(1-\delta)\left(\frac{t}{t_s} \right)^{m-1}+\delta n \left(1-\frac{t}{t_s}\right)^{n-1}}{\delta+(1-\delta)\left(\frac{t}{t_s} \right)^m-\delta\left( 1-\frac{t}{t_s} \right)^n}~~,
\end{equation}
for which eq. (\ref{redshift}) has to be applied.

We consider the scenario in which the universe contains two fluid components, namely non-relativistic matter and the scalar field which drives an exotic singularity. These fluids obey independently their conservation laws. We assume the standard behaviour for the non-relativistic (dust) matter component
\be
\rho_m=\Omega_m\rho_0\left(\frac{a_0}{a}\right)^3
\ee
and the evolution of the other fluid, which we name here $\rho_{\Phi}$, can be determined by taking the difference between whole energy density, $\rho$ as given in  Friedmann eq. (\ref{rho}) and $\rho_m$, i.e.
\be
\rho_{\Phi}=\rho-\rho_m~.
\ee
In fact, it is just the $\rho_{\Phi}$ component of the Universe which is responsible for the appearance of an exotic singularity at $t\rightarrow t_s$.
Using this we can rewrite the Friedmann eq. (\ref{rho}) as
\be
\rho=\frac{3H_0^2}{8\pi G}\left[\Omega_m\left(\frac{a_0}{a}\right)^3+\Omega_{\Phi}\right]
\ee
so that the dark energy density is given by
\be
\Omega_{\Phi}=1-\Omega_{m0}\frac{H_0^2}{H^2}\left(\frac{a_0}{a}\right)^3=1-\Omega_{m}.
\ee
The barotropic index of the equation of state for the dark energy given by the canonical scalar field $\phi$ is defined as $w_{\Phi}=p_{\Phi}/ \rho_{\Phi}$, where $p_{\Phi} = (1/2)\dot{\Phi}^2 - V(\Phi)$ and $\rho_{\Phi} = (1/2)\dot{\Phi}^2 + V(\Phi)$ ($V(\Phi)$ is the potential). In the phantom regime which has negative kinetic energy \cite{phantom} one has $p_{\Phi} = -(1/2)\dot{\Phi}^2 - V(\Phi)$ and $\rho_{\Phi} = -(1/2)\dot{\Phi}^2 + V(\Phi)$. On the other hand, the effective barotropic index of the equation of state is $w_{eff}=p/ \rho$. In the case in which we consider the times when the radiation can be neglected $p=p_{\Phi}$.

The model parameters used here will be the same as the ones taken in our previous paper \cite{lastpaper}; they are listed in Table \ref{tab1}.

\begin{table}
\begin{center}
\begin{tabular}{ >{\centering\arraybackslash}m{1.0in}  >{\centering\arraybackslash}m{.60in} >{\centering\arraybackslash}m{.50in} >{\centering\arraybackslash}m{.50in} >{\centering\arraybackslash}m{.50in} }
\toprule[1.5pt]
{\bf Model} & {\bf m} & {\bf n} &  $\delta$ &  $y_0$  \\
\hline
SFS1        & 2/3   & 1.9999      &  -0.43    &  0.99    \\
SFS2        & 2/3   & 1.9999      &  0        &  0.99     \\
SFS3        & 0.749   & 1.99      &  -0.45   &  0.77    \\
\hline
FSFS1 & 0.56 & 0.8 & 0.42 & 0.96 \\
FSFS2 & 2/3 & 0.7 & 0.0 & 0.79 \\
FSFS3 & 2/3 & 0.7 & 0.24 & 0.96 \\
\bottomrule[1.25pt]
\end{tabular}
\caption{The sets of parameters for the scale factor (\ref{a(t)}) which are used for SFS and FSFS models. See \protect\cite{lastpaper} and references therein for further discussion on these choices.}
\label{tab1}
\end{center}
\end{table}

Note that SFS2 and FSFS2 correspond to the dust limit of these models. Clearly they are amply ruled out, but they provide pedagogically useful fiducial comparisons for some of the discussion that follows.

\begin{figure}
\includegraphics[width=3.5in,keepaspectratio]{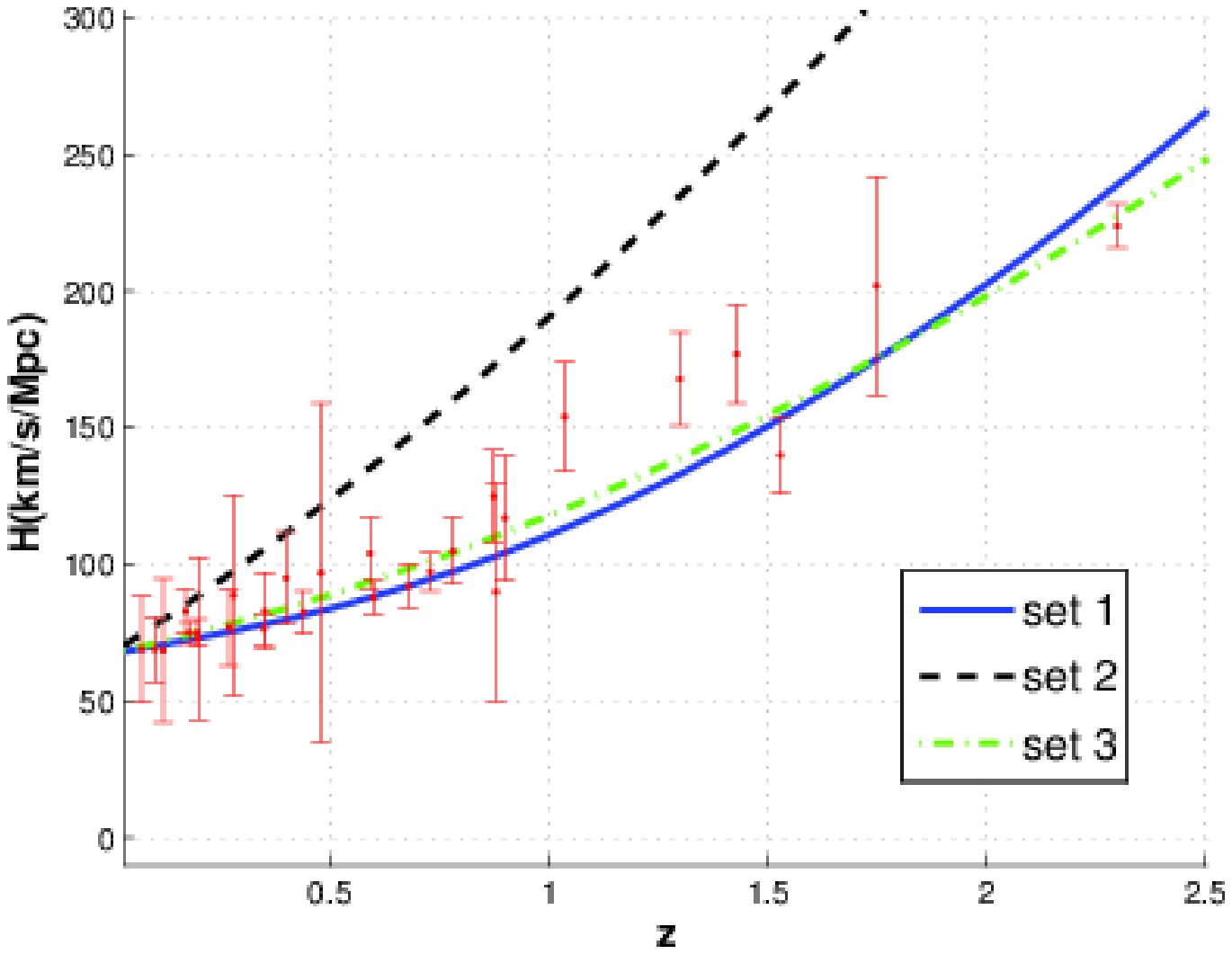}
\includegraphics[width=3.5in,keepaspectratio]{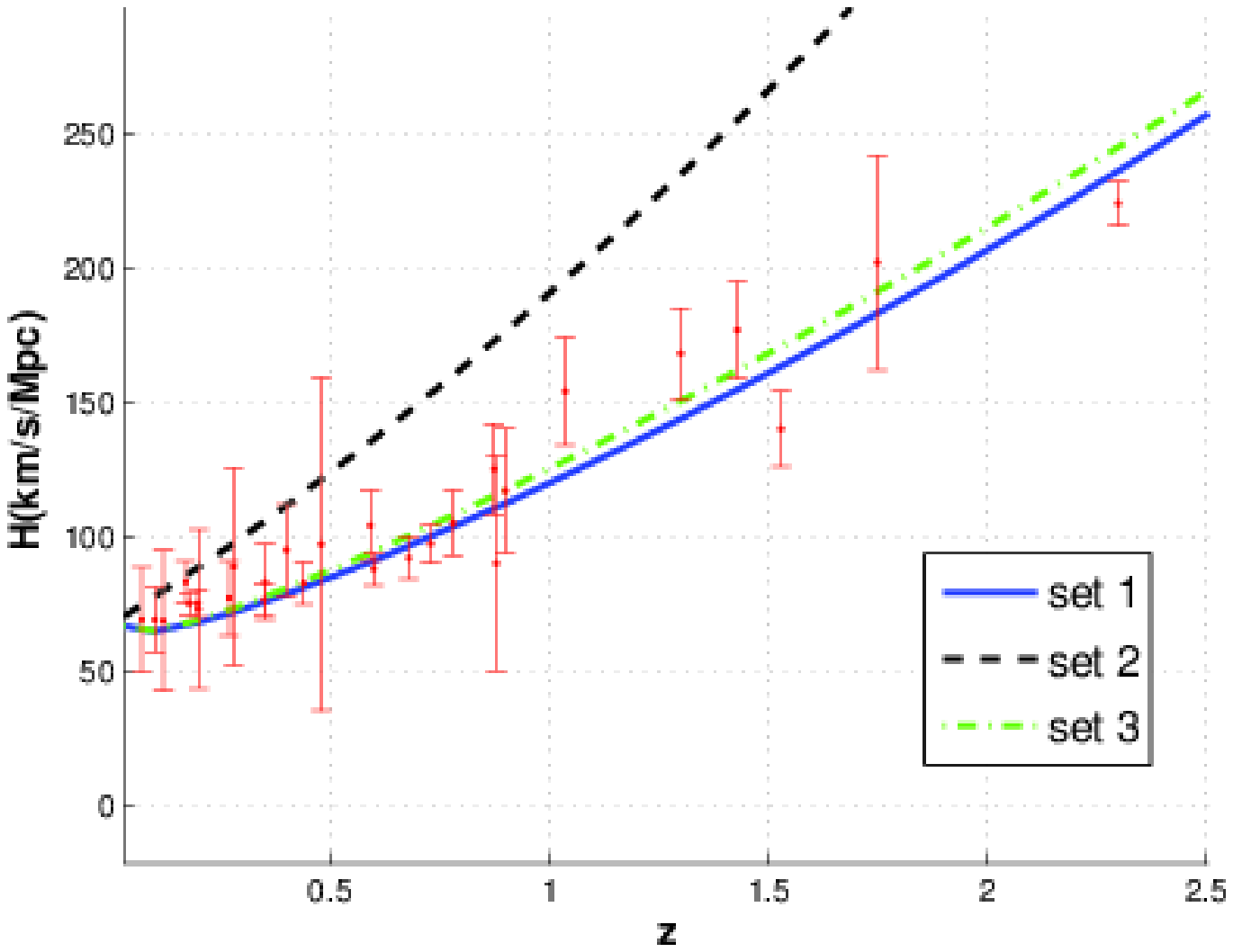}
\caption{The redshift evolution of the Hubble function (\ref{eq:Hsing}) for SFS (top) and FSFS (bottom) type of singularities with the set of parameters shown in Table \ref{tab1}, plotted against the observational data of Ref. \cite{Hmeasurements}.}
\label{fig1}
\end{figure}

Using these parameters for the redshift function (\ref{redshift}) one can check whether our models are consistent with current observations of the Hubble parameter as a function of redshift (\ref{eq:Hsing}) and the plots for our choices of SFS and FSFS parameters given in the Table \ref{tab1} are shown in Fig. (\ref{fig1}), with the observational data taken from the recent compilation \cite{Hmeasurements}. These illustrate the point that the background evolution of the dust models is highly discrepant.

\section{Varying fine-structure constant}
\label{sec:varalpha}

High-resolution spectroscopic observations of absorption clouds along the line of sight of quasars have provided indications of spacetime variations of the fine-structure constant $\alpha$ at the level of a few parts per million, in the approximate redshift range $1<z<4$, the most recent one being that of Ref. \cite{webb}. A possible cause for concern is that these measurements come from archival data, and thus several efforts have been made to independently confirm this result through dedicated measurements. A summary list of some of these new measurements is provided in Table \ref{tab3}; the latest of these efforts is the ongoing Large Program at the VLT UVES (Very Large Telescope Ultraviolet and Visual Echelle Spectrograph) \cite{bonifacio}. We will use both the data in the Table \ref{tab3} and that of Ref. \cite{webb} as sets of data to constrain our models. Note that the former has fewer data points (and a smaller redshift sampling) but smaller uncertainties, the reverse
  being true for the latter.

\begin{table}
\begin{tabular}{|c|c|c|c|c|}
\hline
 Object & z & ${ \Delta\alpha}/{\alpha}$ & Spectrograph & Ref. \\
\hline\hline
HE0515$-$4414 & 1.15 & $-0.1\pm1.8$ & UVES & \protect\cite{alphaMolaro} \\
\hline
HE0515$-$4414 & 1.15 & $0.5\pm2.4$ & HARPS/UVES & \protect\cite{alphaChand} \\
\hline
HE0001$-$2340 & 1.58 & $-1.5\pm2.6$ &  UVES & \protect\cite{alphaAgafonova}\\
\hline
HE2217$-$2818 & 1.69 & $1.3\pm2.6$ &  UVES--LP & \protect\cite{LP1}\\
\hline
Q1101$-$264 & 1.84 & $5.7\pm2.7$ &  UVES & \protect\cite{alphaMolaro}\\
\hline
\end{tabular}
\caption{Currently specific measurements of $\alpha$. The columns respectively contain the object along each line of sight, the redshift of the absorber, the measured variation of fine structure constant $\alpha$ (in parts per million), the name of the spectrograph, and the reference reporting the measurement. The fourth entry corresponds to the recent Large Program measurement.}
\label{tab3}
\end{table}

Any dynamical scalar field providing the dark energy is naturally expected to couple to the rest of the model and in particular to lead to spacetime variations of fundamental couplings \cite{carroll}. The coupling between the scalar-field and the electromagnetic field can be described by
\begin{equation}
{\cal L}_{\Phi, F}=-\frac{1}{4}B_F(\Phi)F_{\mu\nu}F^{\mu\nu}\,,
\end{equation}
where as usual the gauge kinetic function is such that $B_F(\Phi)=\alpha_0/\alpha(\Phi)$. To a good approximation (at least for the low redshifts of interest in the present work) we may assume a linearized gauge kinetic function
\begin{equation}
B_F(\Phi)=1-\xi\kappa(\Phi-\Phi_0)~~,
\end{equation}
where $\kappa^2=8\pi G/c^4$ and $\xi$ parametrizes the coupling between the scalar field and the electromagnetic sector. It then follows that the evolution of $\alpha$ can be written as
\begin{equation}
\frac{\Delta\alpha}{\alpha}\equiv\frac{\alpha-\alpha_0}{\alpha_0}= B_F^{-1}(\Phi)-1=\xi\kappa(\Phi-\Phi_0)~~.
\end{equation}

If one assumes that the same degree of freedom provides all of the dark energy and the variation of $\alpha$, then the dark energy equation of state can be inferred from the dynamics of the field, as first discussed in \cite{nunes}. Using the fact that for a canonical scalar field $\dot{\Phi}^2 = p_{\Phi} + \rho_{\Phi}$ and changing the derivative with respect to time into the derivative with respect to logarithm of the scale factor i.e. that $(...)' \equiv d/d\ln{a} = H^{-1}d/dt$ we have for the dynamics of the scalar field
\be
w_{\Phi}+1 = \frac{\dot{\Phi}^2}{\rho_{\Phi}} = \frac{(\kappa\Phi')^2}{3\Omega_\Phi}~,
\ee
where $\Omega_{\Phi}$ if the fraction of the universe's energy in the scalar field component
\be
\Omega_{\Phi} = \frac{\rho_{\Phi}}{\rho_{\Phi}+\rho_m} = \frac{\rho_{\Phi} a^3}{\rho_0 \Omega_{m0} + \rho_{\Phi} a^3}~~.
\ee
The equation for the field can easily be integrated with respect to the scale factor \cite{calabrese1,pauline1}, and changing variables using $dz/(1+z)=da/a$ we finally find, in terms of the redshfit
\begin{equation}
\label{eq:alphavar1}
\frac{\Delta\alpha}{\alpha}(z)=\xi\int_0^{z}\sqrt{3\Omega_{\Phi}(\hat{z})\mid (1+w(\hat{z}))\mid }\frac{d\hat{z}}{(1+\hat{z})}\,.
\end{equation}

Notice that the above expression is only valid for canonical (quintessence-type) scalar fields. On the other hand, in the phantom regime $w<-1$ (negative kinetic term of the scalar field) we instead have \cite{pauline2}
\begin{equation}
w+1= -\frac{(\kappa\Phi')^2}{3\Omega_\Phi}
\end{equation}
and this now leads to
\begin{equation}\label{eq:alphavar2}
\frac{\Delta\alpha}{\alpha}(z)=-\xi\int_0^{z}\sqrt{3\Omega_{\Phi}(\hat{z})\mid 1+w(\hat{z})\mid }\frac{d\hat{z}}{(1+\hat{z})}\,;
\end{equation}
the extra minus sign comes from the fact that in the canonical case one physically expects the field to be rolling down the potential, while in the phantom case it should be nominally climbing up.

In the above formulas $\Omega_\Phi(z)$ and $w(z)$ are the fraction of the universe's energy in the form of dark energy and its equation of state respectively. We thus see that knowledge of these parameters is sufficient (up to a normalization provided by the coupling $\xi$) to determine the evolution of $\alpha$. Thus with the above assumptions we can easily determine this evolution in the exotic singularity models under consideration.

Note that in some of these models $w(z)$ can cross the $w=-1$ phantom divide. Thus in these models the evolution of $\alpha$ need not be monotonic, but may have inflection points and change sign. On the other hand, this cannot happen in the dust case. This is one reason for keeping this model in the analysis, as a simple comparison point.

\begin{figure}
\includegraphics[width=3.5in,keepaspectratio]{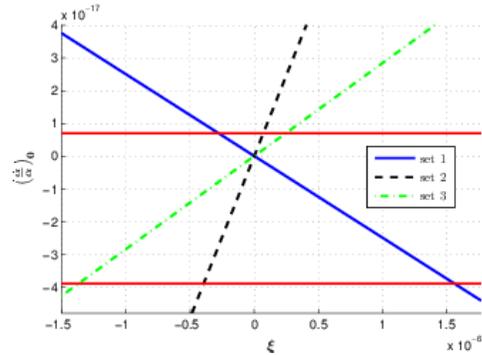}
\caption{The present-day drift rate of the fine-structure constant $\alpha$ as a function of the coupling $\xi$, for the three SFS models under consideration, compared to the one-sigma experimental bound of \protect\cite{Rosenband}.}
\label{fig2}
\end{figure}

In particular the above equations apply at redshift $z=0$, for which atomic clock measurements provide a very tight limit \cite{Rosenband} on the current drift rate of $\alpha$, namely
\begin{equation}\label{rosenb}
\left(\frac{\dot\alpha}{\alpha}\right)_0=(-1.6\pm2.3)\times 10^{-17}{\rm yr}^{-1}\,.
\end{equation}
This bound was later refined (under plausible theoretical assumptions) in \cite{barshaw}, but we use the direct (and more conservative) bound in our analysis. Similarly we do not use Oklo bound \cite{Oklo} at $z = 0.14$: although nominally quite strong, it is subject to much larger theoretical and systematic uncertainties than the spectroscopic measurements we are considering. With the assumptions we are making for this class of models we therefore have from (\ref{eq:alphavar1}) that \cite{pauline1}
\begin{equation}\label{drift0}
\left|\frac{\dot\alpha}{\alpha}\right|_0=|\xi|H_0\sqrt{3\Omega_{\Phi0}\mid 1+w_{\Phi0}\mid }\,,
\end{equation}
where the modulus signs allow for the fact that the models can be at either side of the phantom divide and the sign of the coupling in the gauge kinetic function is not a priori defined. Using the current value of the Hubble constant (say the $H_0=(67.4\pm1.4)\, {\rm km.s}^{-1}{\rm Mpc}^{-1}$ Planck value) one gets to the following conservative ($3\sigma$) bound
\begin{equation}\label{driftbound}
|\xi|\sqrt{3\Omega_{\Phi0}\mid 1+w_{\Phi0}\mid }<10^{-6}\,.
\end{equation}
(Obviously the choice of a different value of $H_0$---say from local measurements---has a negligible effect on the above bound.)
Therefore the different models being considered will be subject to different bounds on $\xi$, since they will have different values of $\Omega_{\Phi0}$ and $w_{\Phi0}$. These bounds are summarized in Table \ref{tab4}, together with the maximum variation of $\alpha$ allowed in each model, up to a redshift $z=5$, when the $\xi$ bound is saturated. The choice of a maximum redshift of $z=5$ is meant to represent the range over which future measurements may be expected, in particular from the European Extremely Large telescope (E-ELT) \cite{eelt}.

\begin{figure*}
\begin{center}
\includegraphics[width=3.5in,keepaspectratio]{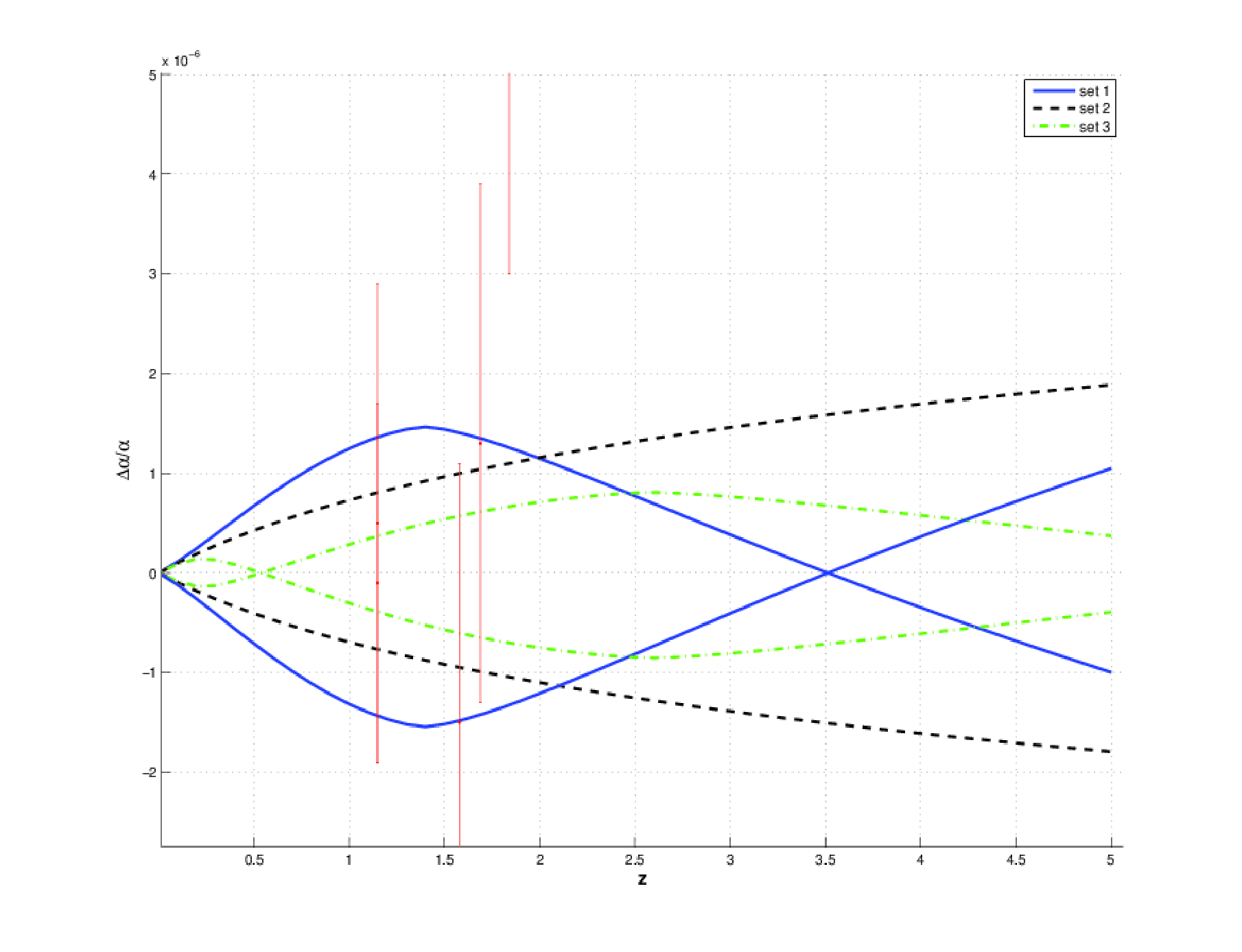}\vskip0in
\includegraphics[width=3.5in,keepaspectratio]{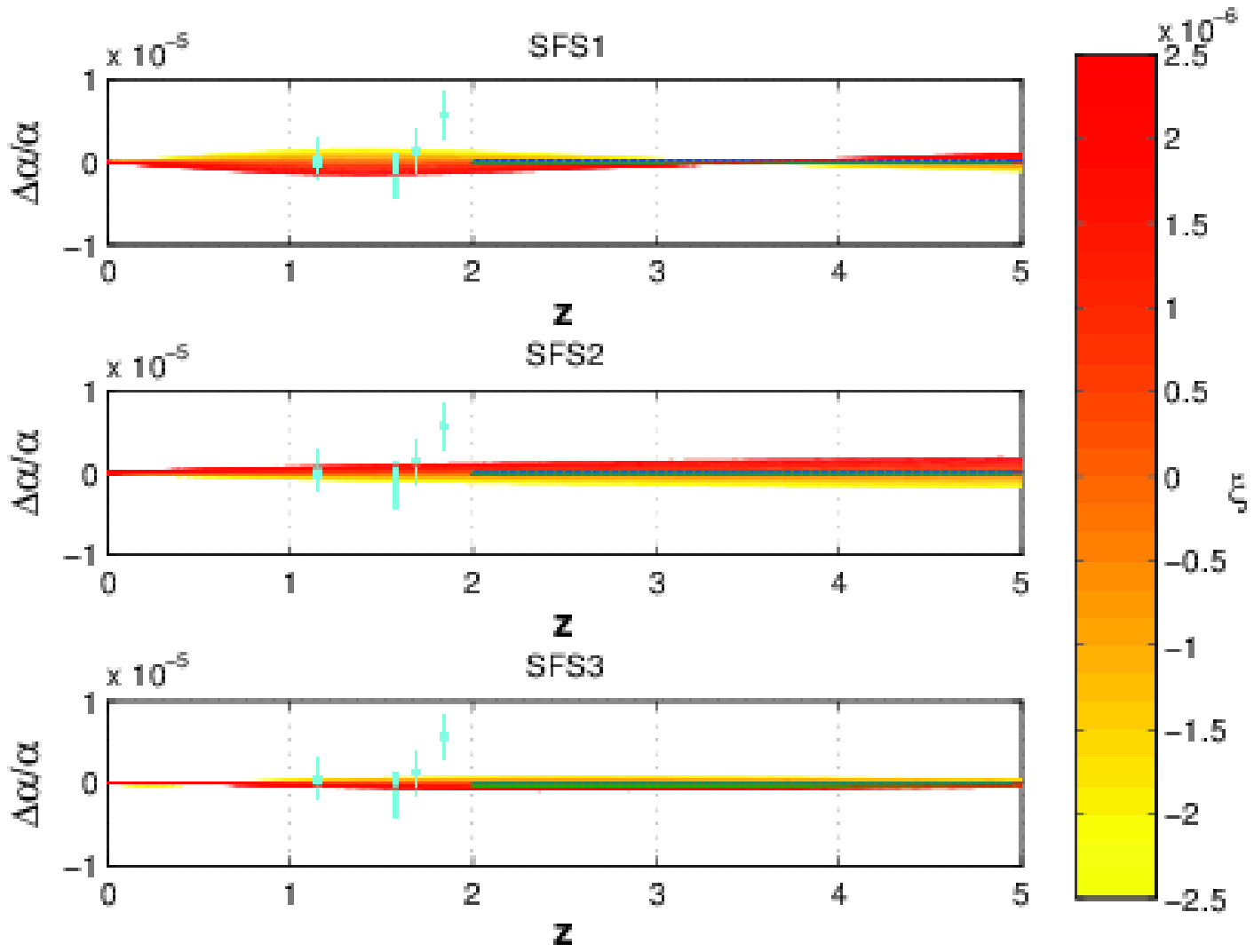}
\includegraphics[width=3.5in,keepaspectratio]{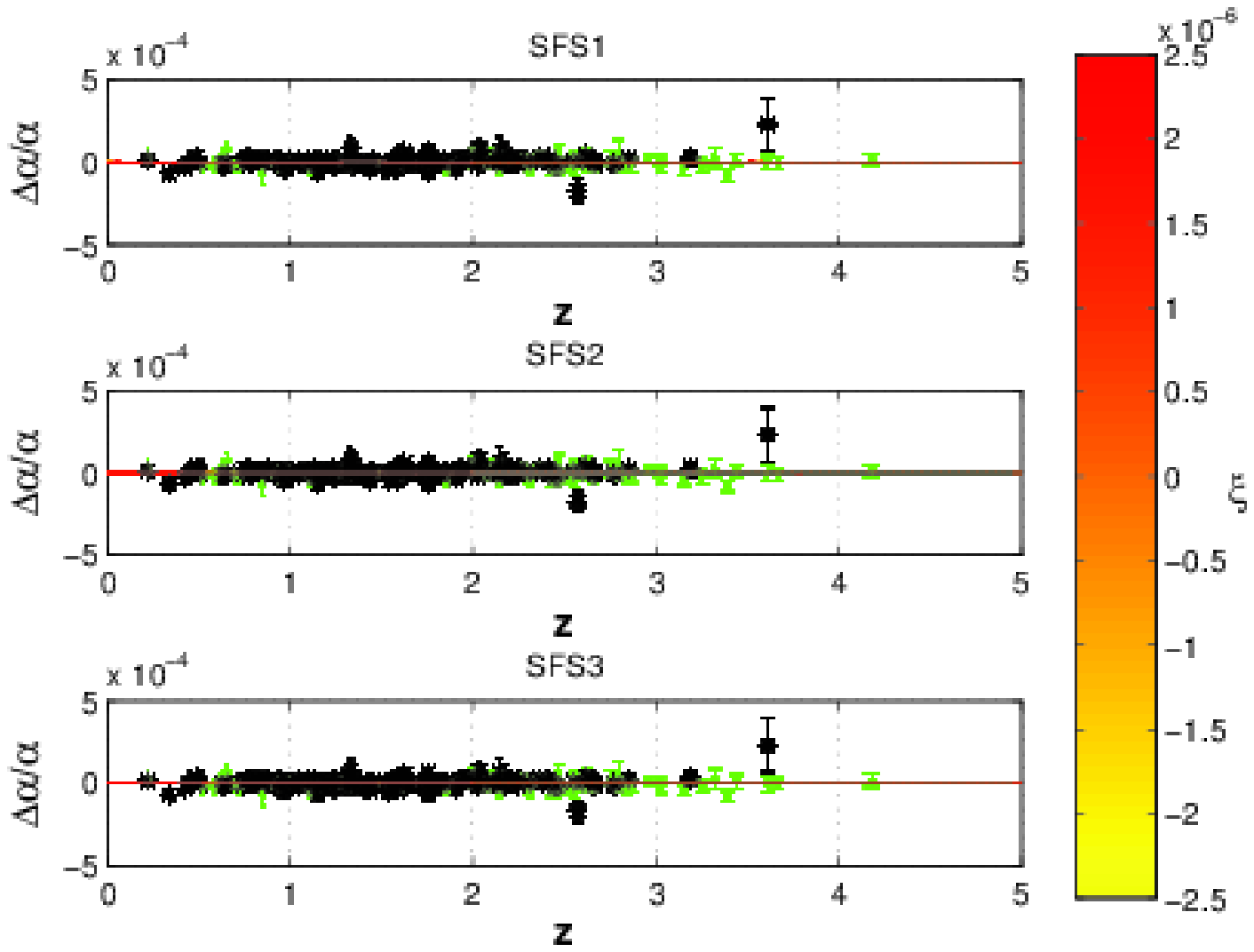}
\end{center}
\caption{The top panel shows the redshift dependence of $\alpha$ for the values of the coupling that saturate the redshift $z=0$ constraints; the bottom panels illustrate the range of allowed variations for each of the models, compared to the dedicated measurements of Table \protect\ref{tab3} and the data of \protect\cite{webb} respectively. The thin black rectangle in the redshift range $2<z<5$ is meant to indicate the expected sensitivity of future E-ELT measurements \protect\cite{hires}.}
\label{fig3}
\end{figure*}

Note that in the case of the FSFS models the extremely negative present-day equation of state leads to a very tight bound on the coupling (coming from atomic clock measurements). The result of this is that the allowed variations of $\alpha$ in these models are extremely small: about two orders of magnitude smaller than would be needed to explain the results of \cite{webb}, and even difficult to detect at all with the next generation of observational facilities. For this reason we will not consider the FSFS class of models any further, focusing instead on the more promising SFS one. An illustration of the atomic clocks bound for these SFS models is also in Fig. \ref{fig2}.

\begin{table}
\begin{center}
\begin{tabular}{|c|c|c|c|c|c|}
\hline
Model & $\Omega_{\Phi0}$ & $w_{\Phi0}$ & $|\xi|_{\rm max} \times 10^6$ &
$z|_{\alpha_{max}}$ & $|\Delta\alpha/\alpha|_{\rm max}\times 10^6$ \\
\hline\hline
SFS1  & $0.685$ & $-1.06$ & $2.76$ & $1.4$ & $1.47$ \\
\hline
SFS2  & $0.685$ & $0.0$ & $0.70$ & $5.0$ & $1.79$ \\
\hline
SFS3  & $0.685$ & $-0.92$ & $2.42$ & $2.6$ & $0.80$\\
\hline\hline
FSFS1  & $0.685$ & $-3.49$ & $0.44$ & $0.2$ & $0.08$\\
\hline
FSFS2  & $0.685$ & $0.0$ & $0.70$ &  $5.0$  &  $1.79$ \\
\hline
FSFS3  & $0.685$ & $-3.68$ & $0.43$ & $0.2$ & $0.06$\\
\hline
\end{tabular}
\caption{Bounds on the coupling $\xi$, coming from the atomic clock measurements of \protect\cite{Rosenband}, for the different models under consideration. Also listed is the maximum allowed variation of $\alpha$, in the redshift range $0 < z \leq 5$, and the redshift at which it occurs, when this bound is saturated. For the dust models the maximum redshift is $z=5$ since the evolution of $\alpha$ is monotonic.}
\label{tab4}
\end{center}
\end{table}

Again we emphasize that while for the dust (SFS2) model the evolution of $\alpha$ is monotonic (and therefore the maximum variation occurs for the highest redshift considered), this is not the case for SFS1 and SFS3. The reason for this is the previously mentioned fact that the dark energy equation of state of these models crosses the phantom divide at some points, the precise redshift of which depends on the choice of model parameters. This can be seen in Fig. \ref{fig3}, which shows the range of allowed variations of $\alpha$ in these models. The bottom two panels of this figure also provide (through the thin black rectangle) a simple visual illustration of the expected sensitivity and redshift span of E-ELT measurements (through the ELT-HIRES instrument \cite{hires}) as compared to currently available measurements.

\section{Results}\label{sec:results}

\begin{figure*}
\includegraphics[width=3.5in,keepaspectratio]{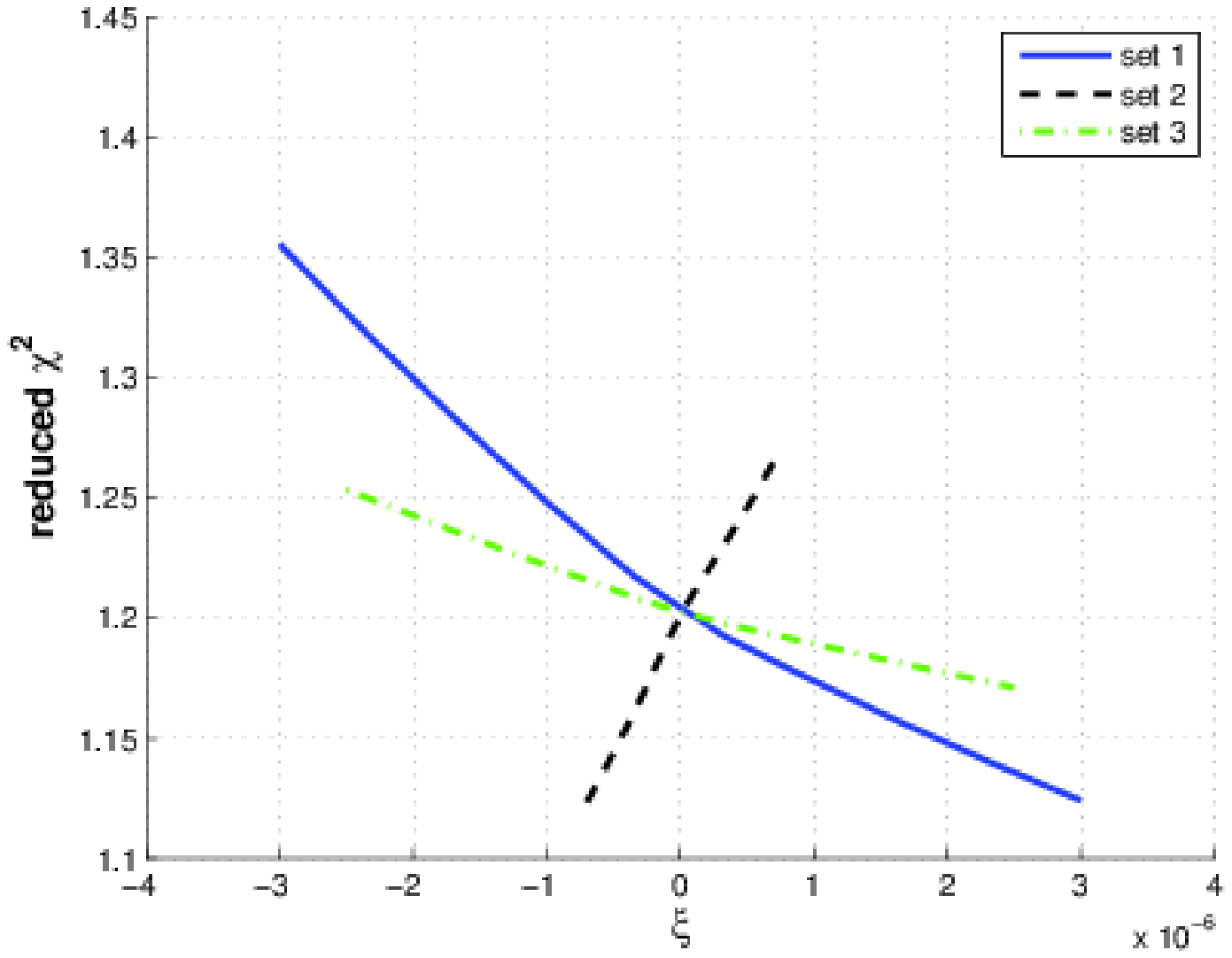}
\includegraphics[width=3.5in,keepaspectratio]{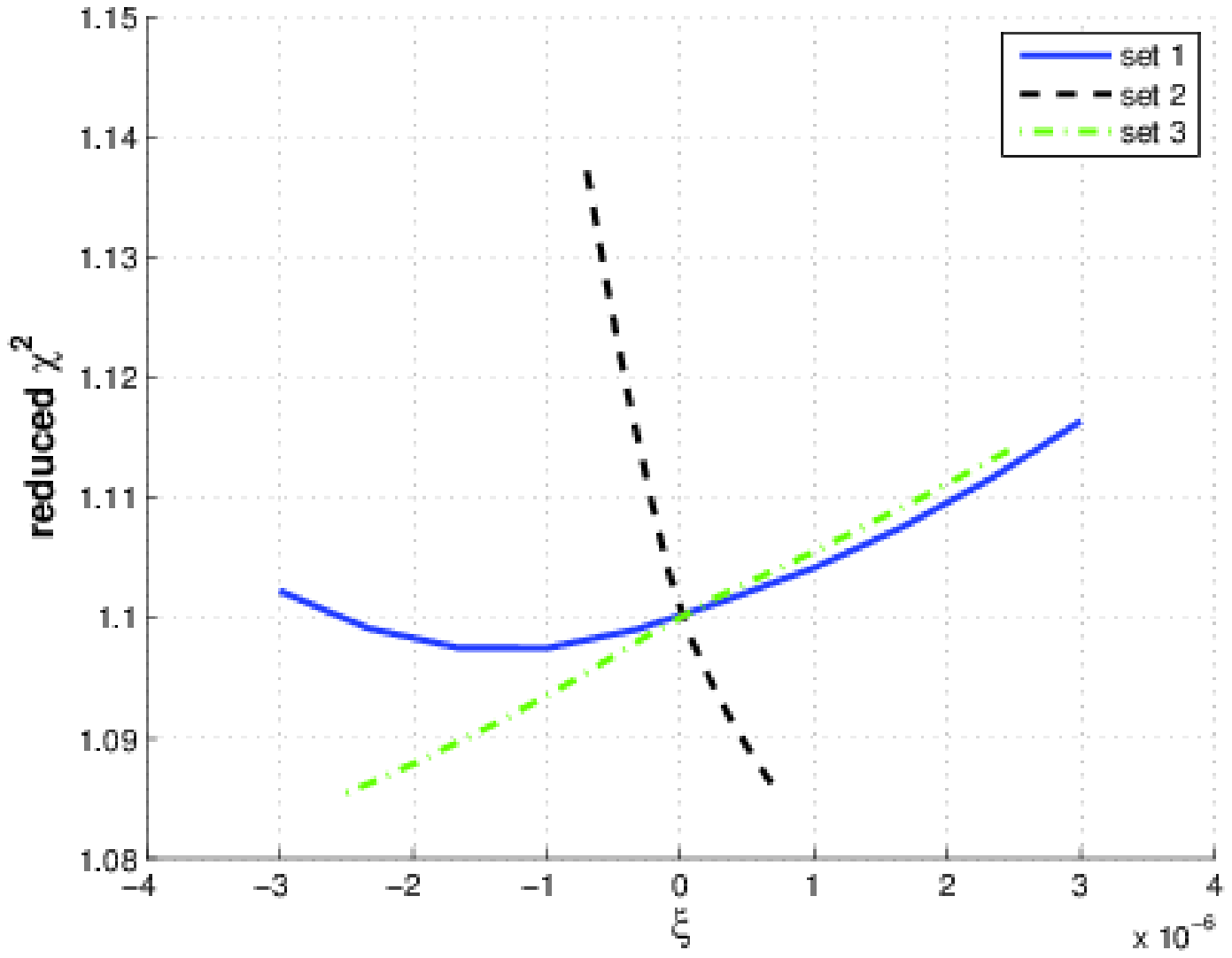}
\includegraphics[width=3.5in,keepaspectratio]{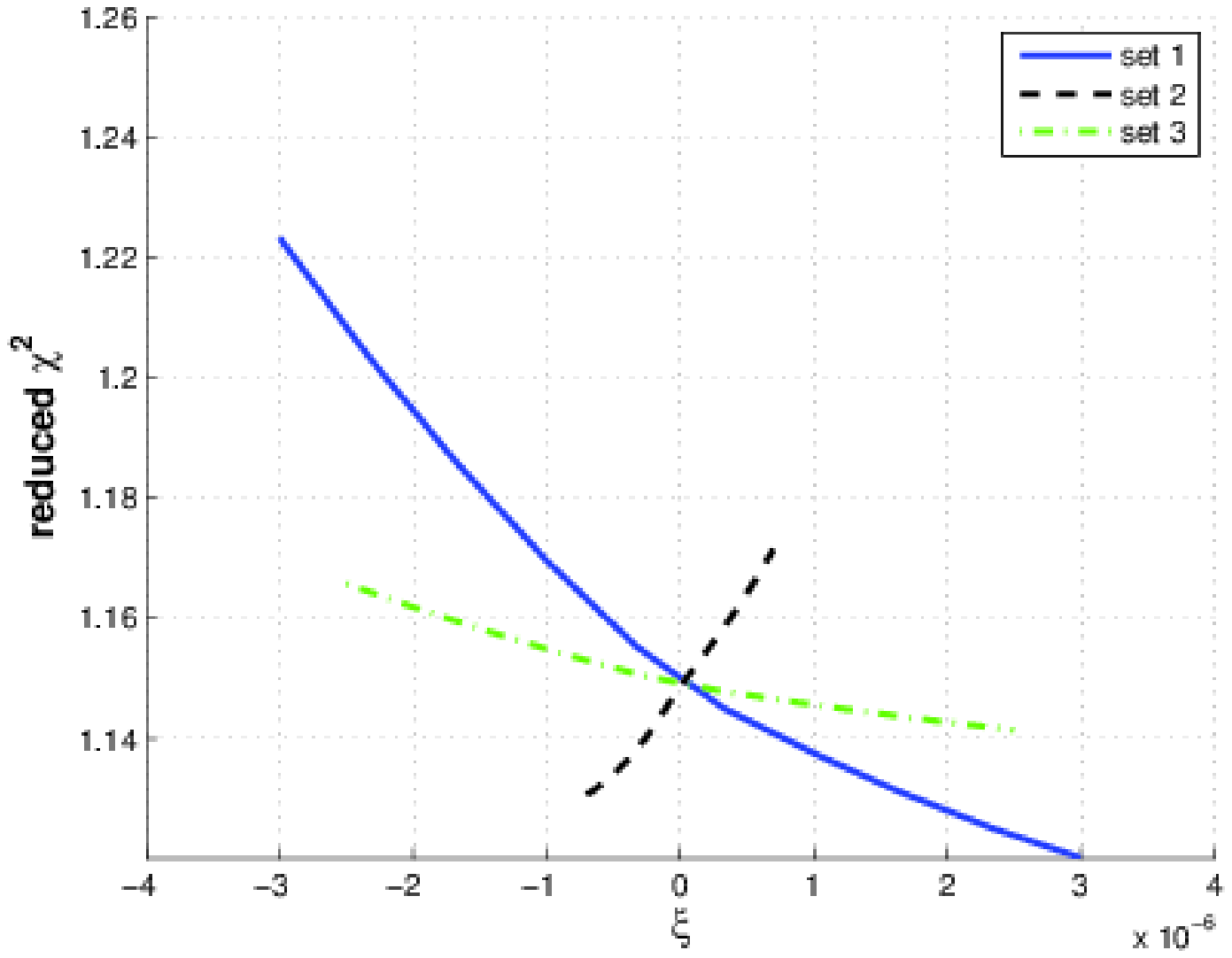}
\includegraphics[width=3.5in,keepaspectratio]{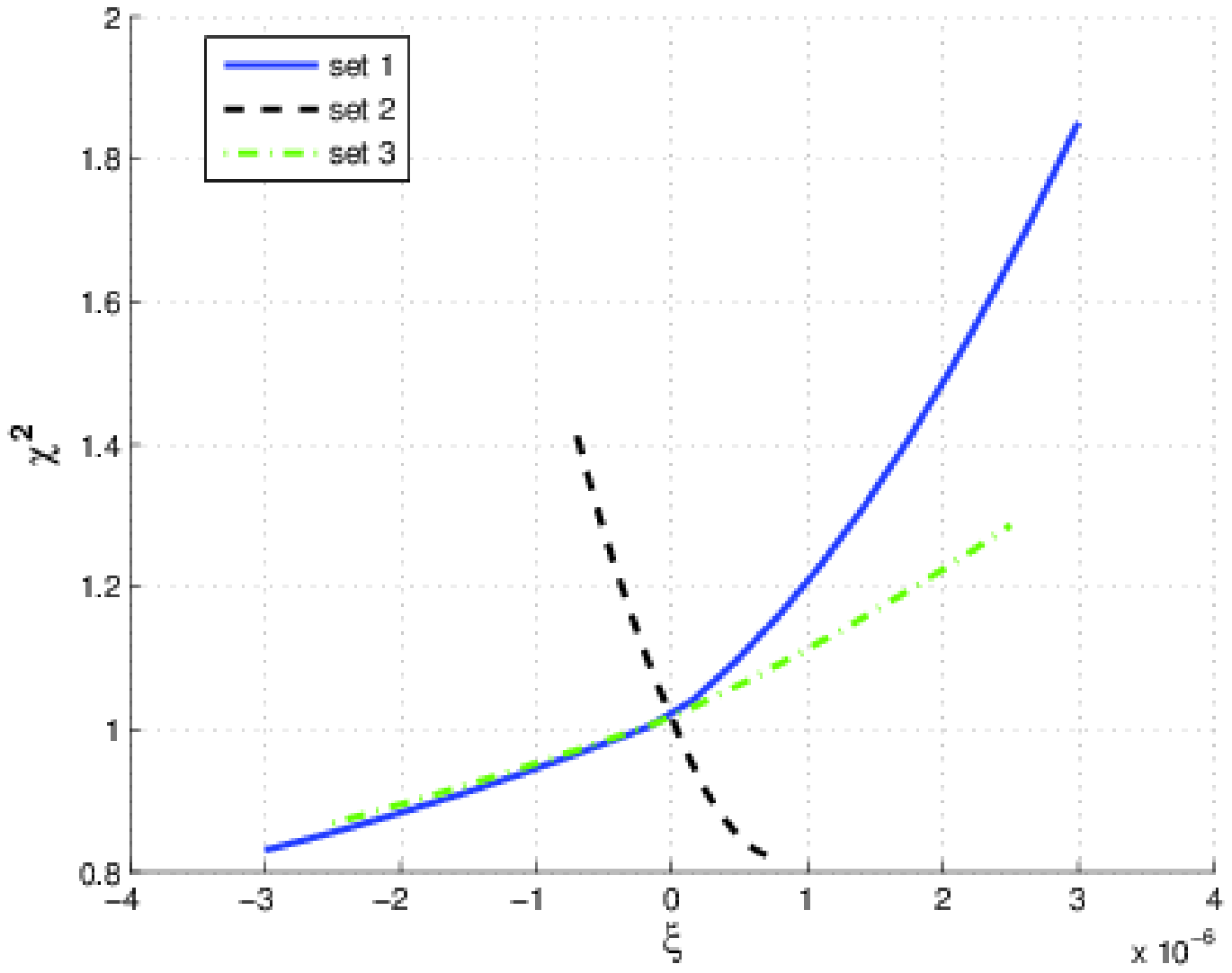}
\caption{Reduced $\chi^2$ for the three SFS models compared with spectroscopic measurements of $\alpha$, in the ranges of the coupling $\xi$ compatibl with local atomic clock bounds. The panels correspond to different datasets: Webb's Keck data (top left), Webb's VLT data (top right), Webb's full dataset (bottom left) and the data in Table \protect\ref{tab3}.}
\label{fig4}
\end{figure*}

We can now compare the SFS models with the spectroscopic measurements of the fine-structure constant $\alpha$ discussed in the previous section, using the standard chi-square statistic.  Figure \ref{fig4} summarizes the results of this comparison, for various choices of dataset: we considered both the dedicated measurements listed in Table \ref{tab3} and the larger archival dataset of Webb \textit{et al.} \cite{webb}, separately including the two subsets of the latter (corresponding to measurements with the Keck and VLT telescopes). For each model we only explore tha range of couplings allowed by the (conservative) bound coming from local atomic clock measurements.

For the dust (SFS2) model, where the $\alpha$ evolution is monotonic, one recovers as expected that the Keck data (which contains predominantly negative measurements) prefers a negative coupling $\xi$, while the VLT data (and also that of Table \ref{tab3}) prefers a positive coupling. The trend is opposite for the SFS1 and SFS3 models, since (as in clear from Fig. \ref{fig3}) the putative underlying scalar field is in the phantom regime for at least part of the redshift range under consideration.

More importantly, one also notices that there is no minimum of the reduced chi-square for this range of couplings. In other words, tha value of coupling that would provide the best fit to any of these spectroscopic datasets would be incompatible with the local atomic clock bound at least at the three sigma level.

\begin{figure}
\includegraphics[width=3.5in,keepaspectratio]{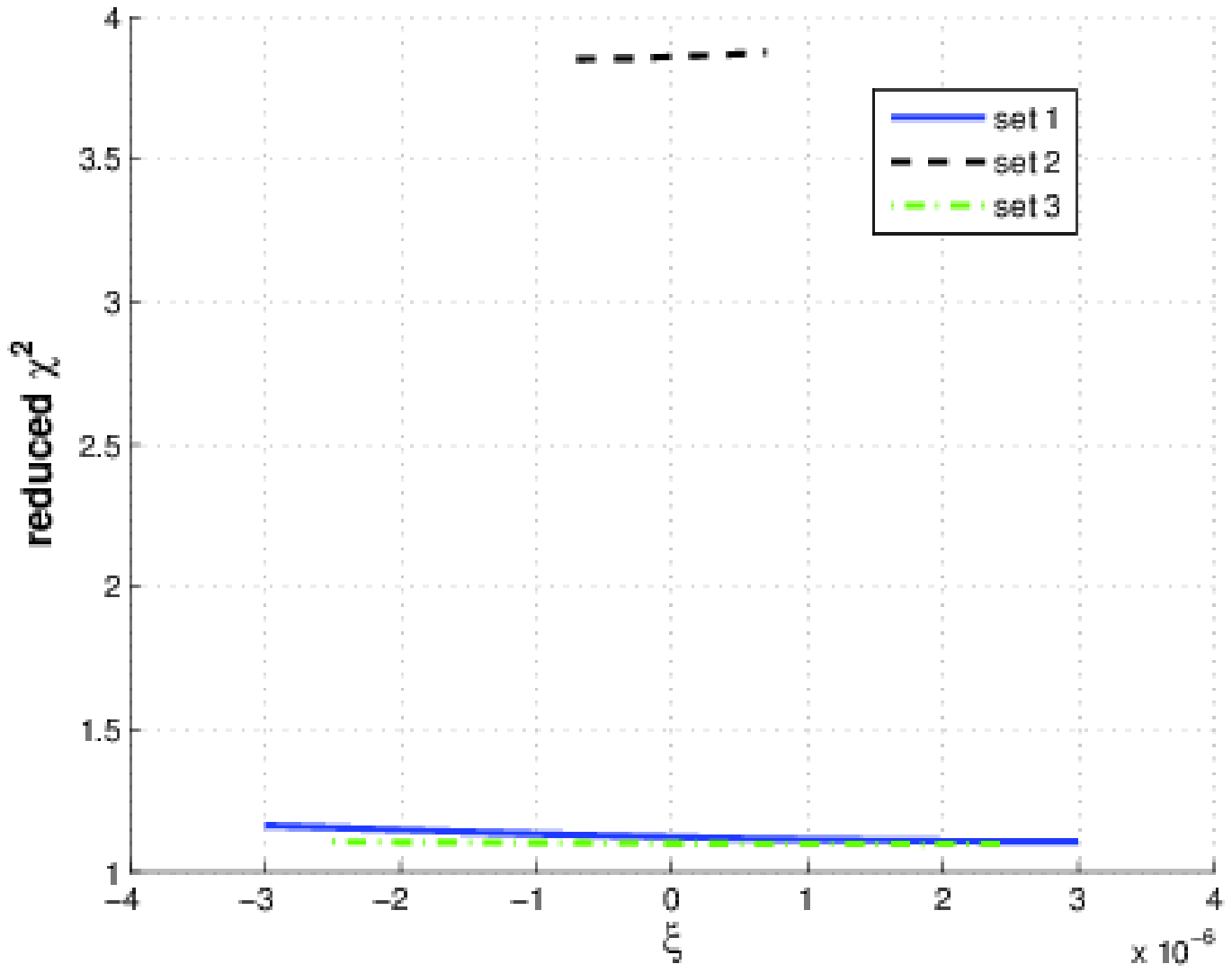}
\includegraphics[width=3.5in,keepaspectratio]{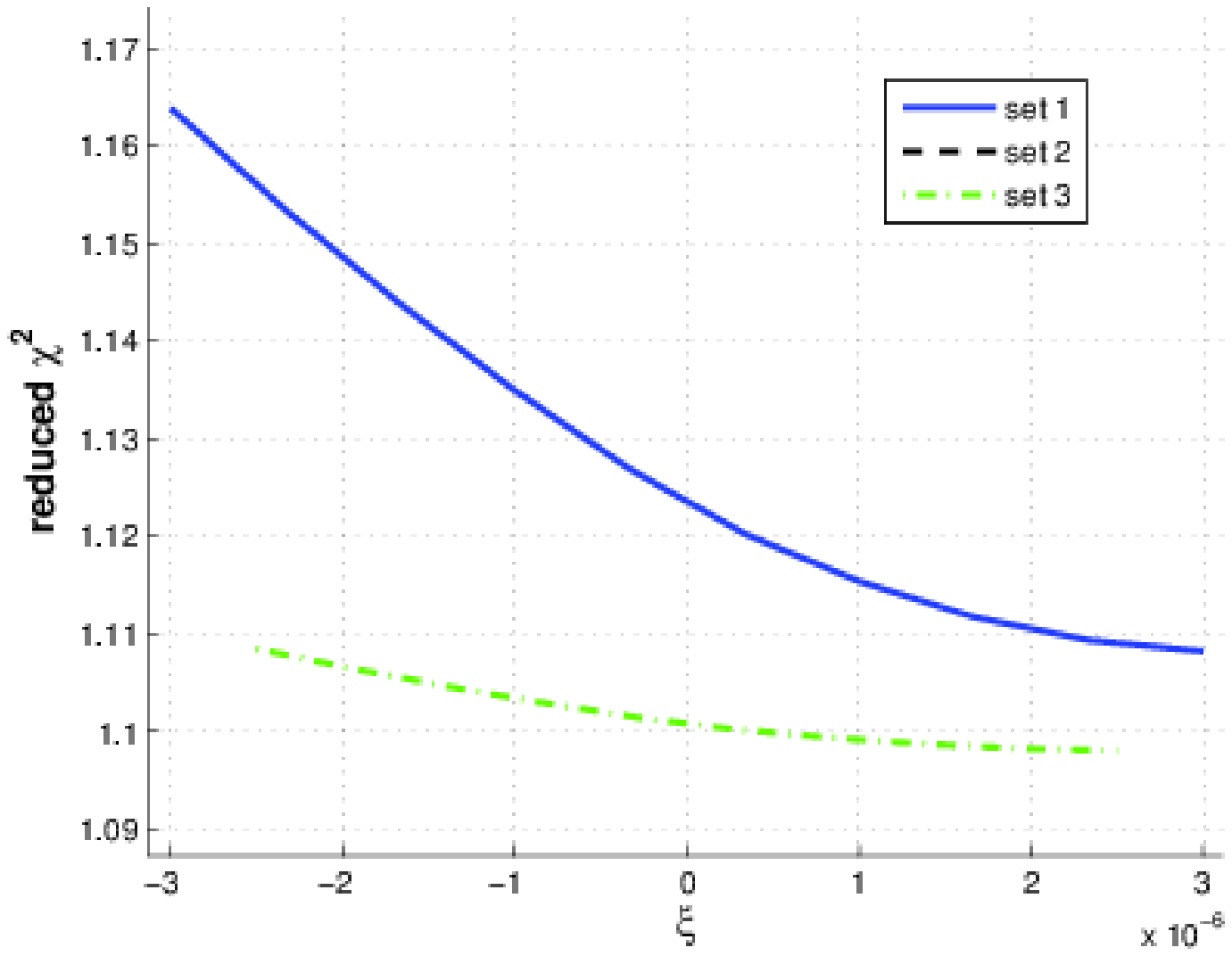}
\caption{Reduced $\chi^2$ for SFS models, using all the $\alpha$ and $H(z)$ data; the bottom panel shows a zoom-in on the SFS1 and SFS3 models (neglecting the dust model).}
\label{fig5}
\end{figure}

A similar analysis can be done considering all the available fine-structure constant measurements as well as the Hubble parameter measurements in \cite{Hmeasurements}; these results are shown in Fig. \ref{fig5}. Naturally the dust (SFS2) model provides an extremely poor fit, while the status of the SFS1 and SFS3 models remains as before (as these are in reasonable agreement with the $H(z)$ data). Thus with the chosen values of the cosmological parameters (which we haven't allowed to vary, as they had been found in previous works to provide the best fits to this class of models), and under the previously discussed assumptions, we find that these models do not provide good fits to available spectroscopic measurements of the fine-structure constant.

\section{Conclusions}\label{sec:conclusions}

The so-called exotic singularity models have been recently suggested as possible mimic models for the observed recent acceleration of the universe. Here we have treated them as toy models for the behavior of an underlying scalar field and, assuming that this also couples to the electromagnetic sector of the theory (which a scalar field would naturally do, unless a new symmetry is postulated to suppress the coupling), calculated the ensuing behavior of the fine-structure constant $\alpha$.

We have shown that with the above assumptions this question can be answered without explicit knowledge of the dynamics of the putative scalar field: the evolution of the dark energy equation of state and density are sufficient, since exotic singularity models assume that the dynamical effects of the field are phenomenologically encoded in the behavior of the scale factor $a(t)$, given by (\ref{a(t)}). We focused on specific choices of SFS and FSFS model parameters, previously shown to be in reasonable agreement with cosmological observations, and used available laboratory and astrophysical tests of the stability of $\alpha$ to further constrain these models.

Our results highlight the importance of local atomic clock measurements such as those of \cite{Rosenband}, in constraining these cosmological models. Specifically, for the FSFS models we considered, the local constraints on the coupling of the putative scalar field to the electromagnetic sector of the theory are so tight that the allowed variations of $\alpha$ at the redshifts probe by optical/UV measurements would be too low to be detected, not only with current spectroscopic facilities but possibly even with future ones. For the SFS class the allowed variations are larger, but nevertheless the values of the coupling $\xi$ that would provide the best fit to currently available spectroscopic measurements of $\alpha$ are in more than three-sigma tensions with the local atomic clock bound.

Nevertheless, at the phenomenological level the SFS models do have one interesting feature: since they can cross the phantom divide (and often do so more than once, at redshifts determined by the model parameters themselves), they will usually lead to a non-monotonic redshift dependence of $\alpha$. This is in contrast with most other single-field, dilaton-type models where its evolution tends to be monotonic---again the dust model we included in the analysis is a simple example of this. Forthcoming more precise measurements with high-resolution ultra-stable spectrographs such as ESPRESSO and ELT-HIRES will allow a detailed mapping of the allowed redshift dependence of $\alpha$ and provide a definitive test of these models.

\section{Acknowledgements}
The research of M.P.D. and T.D. was financed by the National Science Center Grant DEC-2012/06/A/ST2/00395. C.J.M. and P.E.V. are supported by project PTDC/FIS/111725/2009 from FCT, Portugal. C.J.M. is supported by an FCT Research Professorship, contract reference IF/00064/2012, funded by FCT/MCTES (Portugal) and POPH/FSE (EC).\\

\end{document}